 \documentclass[final,3p,times,twocolumn]{elsarticle}
\usepackage{amssymb}
\biboptions{sort&compress}

\journal{NIM B}

\begin{document}

\begin{frontmatter}



\title{Activation cross-sections of longer lived products of deuteron induced nuclear reactions on ytterbium up to 40 MeV}


\author[1]{F. T\'ark\'anyi}
\author[1]{F. Ditr\'oi\corref{*}}
\author[1]{S. Tak\'acs}
\author[2]{A. Hermanne}
\author[3]{H. Yamazaki} 
\author[3]{M. Baba}
\author[3]{A. Mohammadi} 
\author[4]{A.V. Ignatyuk}
\cortext[*]{Corresponding author: ditroi@atomki.hu}

\address[1]{Institute of Nuclear Research of the Hungarian Academy of Sciences (ATOMKI),  Debrecen, Hungary}
\address[2]{Cyclotron Laboratory, Vrije Universiteit Brussel (VUB), Brussels, Belgium}
\address[3]{Cyclotron Radioisotope Center (CYRIC), Tohoku University, Sendai, Japan}
\address[4]{Institute of Physics and Power Engineering (IPPE), Obninsk, Russia}

\begin{abstract}
In the frame of a systematic study of the activation cross-sections of the deuteron induced nuclear reactions, excitation functions of the $^{nat}$Yb(d,xn)$^{177,173,172mg,171mg,170,169}$Lu, $^{nat}$Yb(d,x)$^{175,169}$Yb and $^{nat}$Yb(d,x)$^{173,172,168,167,165}$Tm reactions are studied up to 40 MeV, a few of them for the first time. Cross-sections were measured with the activation method using a stacked foil irradiation technique and high resolution  $\gamma$-ray spectrometry. The experimental data are analyzed and compared to the results of the theoretical model codes ALICE-IPPE, EMPIRE-II and TALYS. From the measured cross-section data integral production yields were calculated. Applications of the new cross-sections are discussed.
\end{abstract}


\end{frontmatter}

% \linenumbers

\section{Introduction}
\label{1}
Activation cross-sections data of deuteron induced nuclear reactions are important both for application in several fields and for benchmarking description of reactions with different model codes. In connection with several projects it was recognized that a high priority needs to be given to establish a reliable database for deuteron activation data. Literature search shows that the status of the available cross-section data for deuteron induced reactions (especially above 15-20 MeV) is very poor. No systematical study has been done earlier and the cross-section data (except a few well measured monitor and medically important reactions) show large discrepancies.
To meet the requirement of improving the reliability of available data, we started to establish an experimental activation database some years ago, by performing new experiments and a systematical survey of published deuteron induced activation cross-sections up to 50 MeV. The targets were irradiated with external beams of the Debrecen, Brussels, Louvain-la-Neuve, and Sendai cyclotrons. The mostly new, measured excitation functions are compared with the results of different nuclear reaction model codes.
As a continuation of our investigation of deuteron induced activation cross-sections on rare earth targets we performed an irradiation of Yb targets with a 40 MeV incident deuteron beam at the Sendai cyclotron. Due to the experimental circumstances (rather long cooling times before start of activity measurement) this experiment results in excitation functions for longer lived Yb, Lu and Tm radionuclides. We have already earlier experimentally studied most of these activation products in proton and alpha-particle induced nuclear reactions on ytterbium \cite{1,2,3}.

\section{Earlier investigations}
\label{2}
Only a few earlier experimental works have been found dealing with deuteron induced activation products on Yb, mostly at lower energies. 
\begin{itemize}
\item	Nichols et al. investigated the excitation functions of  Yb(d,xn)$^{171}$Lu,$^{172}$Lu,$^{173}$Lu between 13.4 to 29.0 MeV \cite{4}.
\item	A. Hermanne et al  investigated the cross sections for deuteron-induced reactions on Yb and Measured cross-sections between 3 and 20 MeV for  Yb(d,xn)$^{170}$Lu/$^{171}$Lu/$^{172}$Lu/$^{173}$Lu/$^{174}$Lu/$^{177}$Lu, and  Yb(d,xnp)$^{169}$Yb/$^{175}$Yb \cite{5}.
\item	S. Manenti et al. measured the activation cross-sections of Yb(d, xn)$^{169}$Lu/$^{170}$Lu/$^{171}$Lu/$^{172}$Lu/$^{173}$Lu/$^{174}$Lu/$^{176}$Lu/$^{177}$Lu, Yb(d,xnp)$^{169}$Yb/$^{175}$Yb/$^{177}$Yb reactions up to 18.18 MeV \cite{6}.
\item	P.P. Dmitriev et al., in the frame of a systematic study: "Yields of Radioactive Nuclides Formed by Bombardment of a Thick Target with 22-MeV Deuterons" measured the thick target yields on Yb for production of $^{173,174}$Lu \cite{7}. 
Considering the theoretical results on production cross-sections for deuteron induced reactions on different target material we have found also only few works: 
\item	Results calculated by the ALICE-IPPE code \cite{8} can be found in \cite{5}. 
\item	The most recent results of systematic calculations made by the TALYS code \cite{9} can be found in the TENDL 2012 on line library \cite{10}.
\item	Comparison of evaluated and experimental data from ENDF/B-VII.1, TENDL-2011 and EXFOR can be found in the JANIS Book of deuteron-induced cross-sections prepared by N. Soppera et al \cite{11}.
\end{itemize}

\section{Experimental}
\label{3}
Elemental cross-sections on $^{nat}$Yb targets were measured using the activation method and the standard stacked foil irradiation technique combined with high-resolution  $\gamma$-ray spectrometry. The irradiation, the activity measurement and the data evaluation were similar as described in more detail in our recent works \cite{12}.
The excitation functions were measured up to 40 MeV by bombarding $^{nat}$Yb targets with low intensity deuteron beams at the AVF-930 cyclotron of the Tohoku University (Sendai). Reactions induced on Al foils present in the stack were used to monitor the parameters of the bombarding deuteron beams. 
The ytterbium (thickness 23 $\mu$m) and aluminum foils (thickness 100 $\mu$m) were of high purity ($\>$ 99.5\%) and purchased from Goodfellow (UK).
The irradiated stack had a complex structure, consisting of a 9 times repeated sequence of  Rh(12.3 $\mu$m), Ho(25 $\mu$m), Au(10.7 $\mu$m), Yb(23 $\mu$m ), MoRe alloy(50 $\mu$m), CuMnNi alloy(25 $\mu$m) foils and the Al(100 $\mu$m) foils serving as monitor.
In principle (n,$\gamma$) and (n,2n) reactions induced by secondary neutrons can also contribute to the production of the isotopes of Yb. We have checked the possible effect of neutron induced reactions by inserting additional target foils at the end of the stack holder, where complete stopping of the deuterons is expected. In the used irradiation setup the effect of neutrons was found to be negligibly small.
The results for the Mn, Rh, Au and MoRe targets were already published \cite{12,13,14,15} and for Ho and Ni target material will be discussed in separate reports.
The irradiation of the targets was performed in He-gas atmosphere in a water cooled target holder. For determination of the beam intensity and energy, the complete excitation functions of the $^{27}$Al(d,x)$^{22,24}$Na monitor reactions were measured, simultaneously with the excitation functions of reactions induced on Yb targets (see Fig. 1). The target stack was irradiated with a collimated deuteron beam of 40 MeV incident energy, for 30 min at about 24 nA. 

\begin{figure}[h]
\includegraphics[scale=0.3]{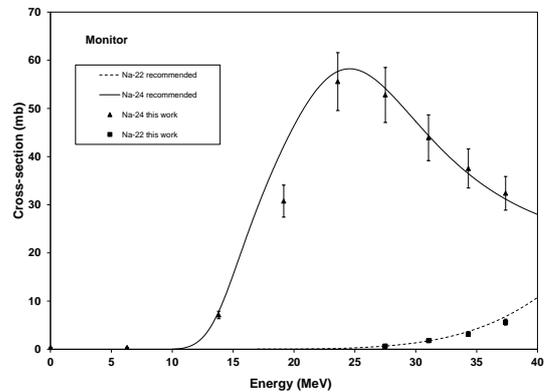}
\caption{Application of monitoring using the $^{27}$Al(d,x)$^{22,24}$Na reaction}
\end{figure}

The radioactivity of each sample and monitor foil was assessed nondestructively by HPGe $\gamma$-spectrometry in two series of measurements. The first counting series was started about 20 hours after the end of the bombardment (EOB) (each foil was measured for 10 min), the second series started about 20 days after EOB (each foil was measured for 5-8 h). The large number of irradiated foils and the high induced activity prevented the optimization of the spectrum measurements of the samples. The information on activity of short-lived radionuclides was lost due to initial one day waiting time after EOB. Due to the large number of foils and the limited number of detectors (2), the measuring time of the first assessment was restricted to 10 min. An additional inconvenience of these circumstances is that detection of the weak  $\gamma$-lines of medium half-life radioisotopes was also impossible. 
The decay data were taken from NUDAT 2.6 \cite{16}, the standard cross-section data of the used monitor reactions $^{nat}$Al(p,x)$^{22,24}$Na  from IAEA recommended database \cite{17}. For the reaction Q-values we used the NNDC Q-value calculator \cite{18}. The energy degradation along the stack was determined via calculation \cite{19} and corrected on the basis of the simultaneously measured monitor reactions by the method described in \cite{20}. The excitation functions were hence determined in an accurate way relative to the simultaneously re-measured monitor reactions.
The decay data and the contributing reactions with their Q-values are presented in Table 1.
The naturally occurring ytterbium is composed of seven stable isotopes, therefore, so called elemental cross-sections were determined except for $^{177}$Lu, where only (d,p) and (d,n) reactions on $^{176}$Yb can contribute (see Table 1). 
The uncertainties on the cross-section values were estimated in the standard way: the independent relative errors of the linearly contributing processes were summed quadratically and the square root of the sum was taken \cite{21}. The uncertainty of the energy of the bombarding beam in the middle of each irradiated foil was estimated combining the uncertainties of the energy degradation depending on the thickness of the target foils in the stack and the energy straggling.

\begin{table*}[ht]
\tiny
%\small
\caption{Decay characteristic of the investigated reaction products}
\centering
\begin{center}
\begin{tabular}{|p{0.4in}|p{0.6in}|p{0.7in}|p{0.5in}|p{0.8in}|p{0.8in}|} \hline 
Nuclide & Half-life & E${}_{\gamma}$(keV) & I${}_{\gamma }$(\%) & Contributing reaction & Q-value\newline (keV)\newline  \\ \hline 
${}^{177g}$Lu\newline  & 6.647 d & 112.9498\newline 208.3662 & 6.17\newline 10.36 & ${}^{176}$Yb(d,n)\newline ${}^{177}$Yb decay & 3959.3 \\ \hline 
${}^{173}$Lu & 1.37 y & 78.63\newline 100.724\newline 272.105 & 11.9\newline 5.24\newline 21.2 & ${}^{172}$Yb(d,n)\newline ${}^{173}$Yb(d,2n)\newline ${}^{174}$Yb(d,3n)\newline ${}^{176}$Yb(d,5n) & 2689.9\newline -3677.41\newline -11142.05\newline -23829.22 \\ \hline 
${}^{172g}$Lu & 6.70 d & 78.7426\newline 181.525\newline 810.064\newline 900.724\newline 912.079\newline 1093.63 & 10.6 \newline 20.6 \newline 16.6 \newline 29.8\newline 15.3\newline 63  & ${}^{171}$Yb(d,n)\newline ${}^{172}$Yb(d,2n)\newline ${}^{173}$Yb(d,3n)\newline ${}^{174}$Yb(d,4n)\newline ${}^{176}$Yb(d,6n) & 2493.6\newline -5525.86\newline -11893.17\newline -19357.81\newline -32044.98 \\ \hline 
${}^{171g}$Lu & 8.24 d & 667.422\newline 739.793\newline 780.711\newline 839.961 & 11.1 \newline 47.9 \newline 4.37\newline 3.05 & ${}^{170}$Yb(d,n)\newline ${}^{171}$Yb(d,2n)\newline ${}^{172}$Yb(d,3n)\newline ${}^{173}$Yb(d,4n)\newline ${}^{174}$Yb(d,5n)\newline ${}^{176}$Yb(d,7n) & 2128.98\newline -4485.5\newline -12504.96\newline -18872.29\newline -26336.93\newline -39024.1 \\ \hline 
${}^{170}$Lu & 2.012 d & 84.262\newline 193.13\newline 572.20\newline 985.10\newline 1054.28\newline 1138.65\newline 1280.25\newline 1341.20\newline 1364.60   & 8.7 \newline 2.07 \newline 1.25 \newline 5.4  \newline 4.60\newline 3.49 \newline 7.9 \newline 3.15\newline 4.47 & ${}^{170}$Yb(d,2n)\newline ${}^{171}$Yb(d,3n)\newline ${}^{172}$Yb(d,4n)\newline ${}^{173}$Yb(d,5n)\newline ${}^{174}$Yb(d,6n)\newline  & -6465.7\newline -13080.2\newline -21099.6\newline -27466.9\newline -34931.6\newline  \\ \hline 
${}^{169}$Lu & 32.018 d & 109.77924\newline 130.52293\newline 177.21307\newline 197.95675 & 17.39\newline 11.38\newline 22.28\newline 35.93\newline  & ${}^{168}$Yb(d,n)\newline ${}^{170}$Yb(d,3n)\newline ${}^{171}$Yb(d,4n)\newline ${}^{172}$Yb(d,5n)\newline ${}^{173}$Yb(d,6n) & 1567.07\newline -13769.87\newline -20384.37\newline -28403.84\newline -34771.16 \\ \hline 
${}^{167}$Lu & 51.5 m & 178.87\newline 213.20\newline 239.22\newline 401.17\newline 1267.26 & 2.5 \newline 3.33\newline 7.7 \newline 3.17\newline 3.87    & ${}^{168}$Yb(d,3n)\newline ${}^{170}$Yb(d,5n)\newline ${}^{171}$Yb(d,6n)\newline ${}^{172}$Yb(d,7n) & -15151.7\newline -30488.7\newline -37103.2\newline -45122.7 \\ \hline 
${}^{177}$Yb & 1.911 h & 150.3 & 20.5 & ${}^{176}$Yb(d,p) & 3341.834 \\ \hline 
${}^{175}$Yb & 4.185 d & 113.805\newline 282.522\newline 396.329 & 3.87\newline 6.13 \newline 13.2 & ${}^{174}$Yb(d,p)\newline ${}^{176}$Yb(d,p2n)\newline ${}^{175}$Tm decay & 3597.7842\newline -9089.38 \\ \hline 
${}^{1}$${}^{69}$Yb\newline  & 32.018 d & 109.77924\newline 130.52293\newline 177.21307\newline 197.95675\newline 307.52\newline 307.73586 & 0.1739\newline 0.1138\newline 0.2228\newline 0.3593\newline 0.003\newline 0.1005\newline  & ${}^{168}$Yb(d,p)\newline ${}^{170}$Yb(d,p2n)\newline ${}^{171}$Yb(d,p3n)\newline ${}^{172}$Yb(d,p4n)\newline ${}^{173}$Yb(d,p5n)\newline ${}^{174}$Yb(d,p6n)\newline ${}^{169}$Lu decay & 4642.414\newline -10694.52\newline -17309.02\newline -25328.49\newline -31695.81\newline -39160.45 \\ \hline 
${}^{173}$Tm${}^{ }$ & 8.24 h & 398.9\newline 461.4 & 87.9 \newline 6.9  & ${}^{173}$Yb(d,2p)\newline ${}^{174}$Yb(d,2pn)\newline ${}^{176}$Yb(d,2p3n) & -2739.62\newline -10204.25\newline -22891.43 \\ \hline 
${}^{168}$Tm${}^{  }$ & 93.1 d & 79.804\newline 184.295\newline 198.251\newline 447.515\newline 720.392\newline 741.355\newline 815.989\newline 821.162 & 10.8\newline 17.9\newline 53.8\newline 23.7\newline 12.0\newline 12.6 \newline 50.3\newline 11.8 & ${}^{168}$Yb(d,2p)\newline ${}^{170}$Yb(d,2p2n)\newline ${}^{171}$Yb(d,2p3n)\newline ${}^{172}$Yb(d,2p4n)\newline ${}^{173}$Yb(d,2p5n)\newline  & -1699.2\newline -17036.15\newline -23650.65\newline -31670.12\newline -38037.44 \\ \hline 
${}^{167}$Tm${}^{  }$ & 9.25 d & 207.801\newline 531.54 & 42\newline 1.61 & ${}^{168}$Yb(d,2pn)\newline ${}^{170}$Yb(d,2p3n)\newline ${}^{171}$Yb(d,2p4n)\newline ${}^{172}$Yb(d,2p5n)\newline ${}^{167}$Yb decay & -8539.87\newline -23876.82\newline -30491.32\newline -38510.79\newline  \\ \hline 
${}^{165}$Tm${}^{  }$ & 30.06 h & 242.917\newline 297.369\newline 460.263 & 35.5\newline 12.7\newline 4.12 & ${}^{168}$Yb(d,2p3n)\newline ${}^{170}$Yb(d,2p5n)\newline ${}^{165}$Yb decay & -24294.89\newline -39631.84\newline  \\ \hline 
\end{tabular}
\end{center}
\begin{flushleft}
\footnotesize{\noindent Abundance of isotopes in natural Yb (\%): ${}^{168}$Yb-0.13, ${}^{170}$Yb -3.05, ${}^{171}$Yb -14.3, ${}^{172}$Yb -21.9, ${}^{173}$Yb -16.12, ${}^{174}$Yb-31.8, ${}^{176}$Yb -12.7

\noindent The Q-values refer to formation of the ground state and are obtained from [18]

\noindent When complex particles are emitted instead of individual protons and neutrons the Q-values have to be decreased by the respective binding energies of the compound particles: np-d, +2.2 MeV; 2np-t, +8.48 MeV; n2p-${}^{3}$He, +7.72 MeV; 2n2p-$\alpha$, +28.30 MeV}
\end{flushleft}
\end{table*}

\section{Comparison with the results of model codes}
\label{4}
The experimental data are compared with the cross-section data reported in the last two TALYS based TENDL 2011 and TENDL 2012 Activation Data Libraries \cite{10} to show the agreement with the experimental data and contribute to the development of the TENDL data library. The cross-sections of the investigated reactions were calculated by us,using the modified pre-compound model codes ALICE-IPPE \cite{8} and EMPIRE-II \cite{22}.
In the modified ALICE IPPE-D and EMPIRE-D code versions the direct (d,p) channel is increased strongly, which gives better agreement for the (d,p) channel, but naturally it reflects  also in the results for all other reactions \cite{23,24}.

\section{Results}
\label{5}
The measured experimental cross-section data are shown in Figs 2-15 together with the earlier literature results and results of the theoretical calculations. The numerical values are collected in Tables 2-3. The following notations were used: 
\begin{itemize}
\item	the "mg" denotes the activation cross-section of the ground state after the "complete" decay of the significantly shorter half-life isomeric state decaying partly or completely to that ground state;
\item	the "cum" denotes the activation cross-section of the final product after the "complete" decay of the simultaneously produced significantly shorter half-life  parent nuclei decaying partly or completely to the investigated final product.
\end{itemize}

\subsection{Cross sections of lutetium radioisotopes}
\label{5.1}
Except for production of $^{177}$Lu only direct (d,xn) reactions contribute to the formation of the measured lutetium radioisotopes.

\subsubsection{$^{176}$Yb(d,xn)$^{177}$Lu}
\label{5.1.1}
In this study, only the cumulative production of $^{177g}$Lu (6.71 d half-life) following total decay of parent $^{177}$Yb (T1/2 = 1.9 h) was assessed. It practically contains no contribution from the internal decay of $^{177m}$Lu (160.4 d, IT 21.7\%) (long-lived, low formation cross-section).
The medically relevant $^{177g}$Lu can be produced in a direct reaction via $^{176}$Yb(d,n)$^{177g}$Lu or indirectly by decay of parent $^{177}$Yb obtained through the $^{176}$Yb(d,p)$^{177m+g}$Yb route. According to the earlier low-energy measurements \cite{5,6} and to the results of the model calculations, the (d,p) reaction has a significant predominance over the direct (d,n) process. The cross-section data above 20 MeV obtained in this study, have large uncertainties due to the separation process of the 208 keV  $\gamma$-line of $^{177g}$Lu overlapping with the  $\gamma$-line at nearly the same energy of $^{167}$Tm (with a half-life of 9.25 d, similar to $^{177g}$Lu).
Our new data are higher than the literature experimental data in the overlapping energy region (Fig. 2). The TENDL 2011 values present total cross-sections of $^{177}$Lu but in the TENDL 2012 already the cumulative cross-section of the $^{177g}$Lu can be deduced. Taking into account the small cross-sections of the direct production of the $^{177m}$Lu the TENDL 2011 results reproduce well the experimental data, both the shape and the magnitude. In case of the TENDL 2012 the underestimation of the experimental data is significant.

\begin{figure}[h]
\includegraphics[scale=0.3]{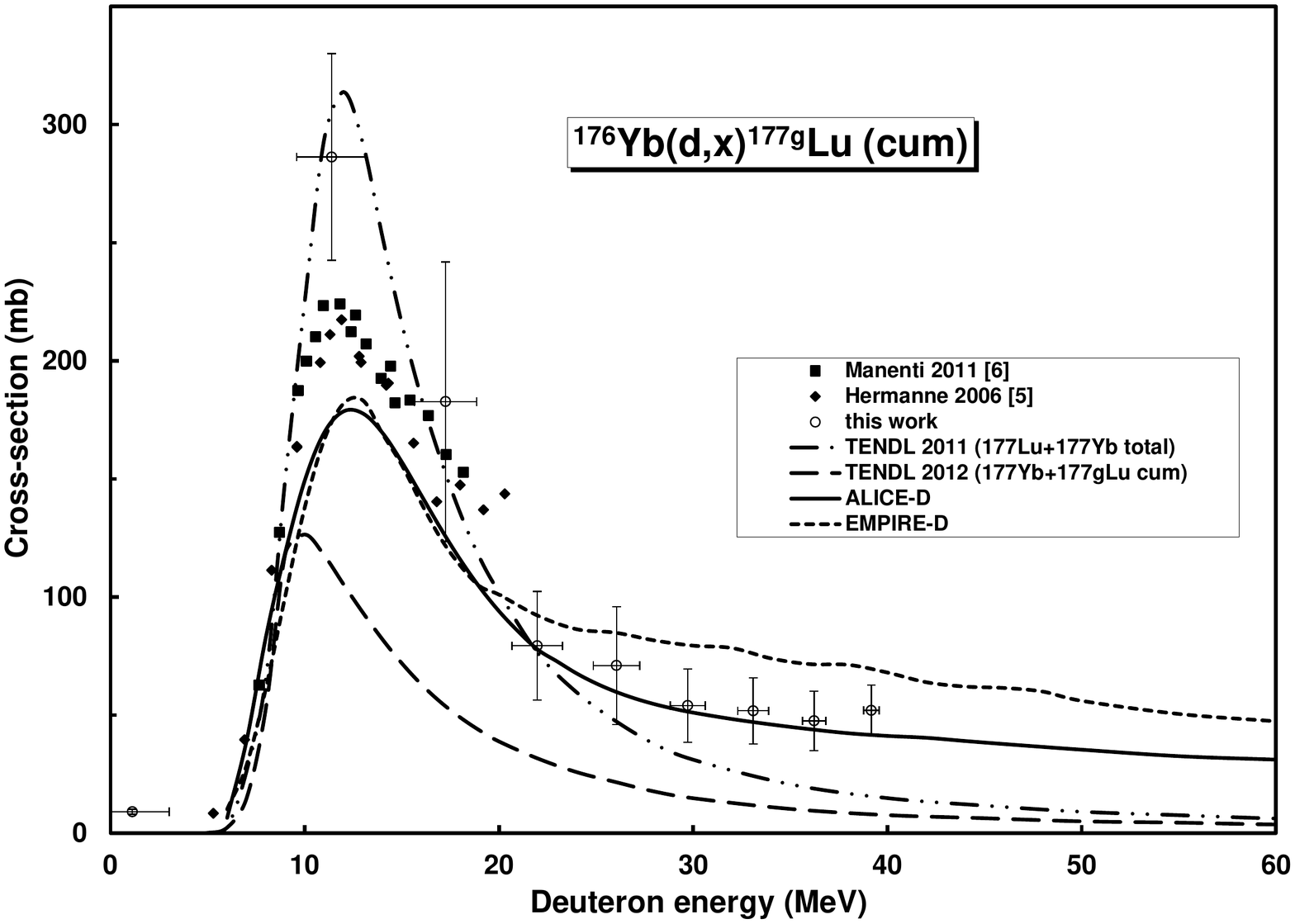}
\caption{Excitation function of the $^{176}$Yb(d,x)$^{177}$Lu process}
\end{figure}

\subsubsection{$^{nat}$Yb(d,xn)$^{173}$Lu}
\label{5.1.2}
The production of $^{173}$Lu (1.37 a) arises from reactions on four stable Yb isotopes. The contribution of the $^{172}$Yb(d,n) reaction is small and cannot be distinguished on Fig. 3. The $^{173}$Yb(d,2n) and $^{174}$Yb(d,3n) produce the first maximum and the $^{176}$Yb(d,5n) reaction the second one (Fig. 3). The agreement with the earlier experimental data of Hermanne et al. \cite{5} and Manenti et al. \cite{6} is acceptable good, but there is significant disagreement with the results of Nichols et al. \cite{4}. It is difficult to judge the agreement with the TENDL predictions: at the low energy maximum it seems to be good, but at higher energies the experimental data are higher. ALICE-D and EMPIRE-D runs together with the TALYS results up to 15 MeV, at the maximum both give higher values, and at higher energies give similar results again.

\begin{figure}[h]
\includegraphics[scale=0.3]{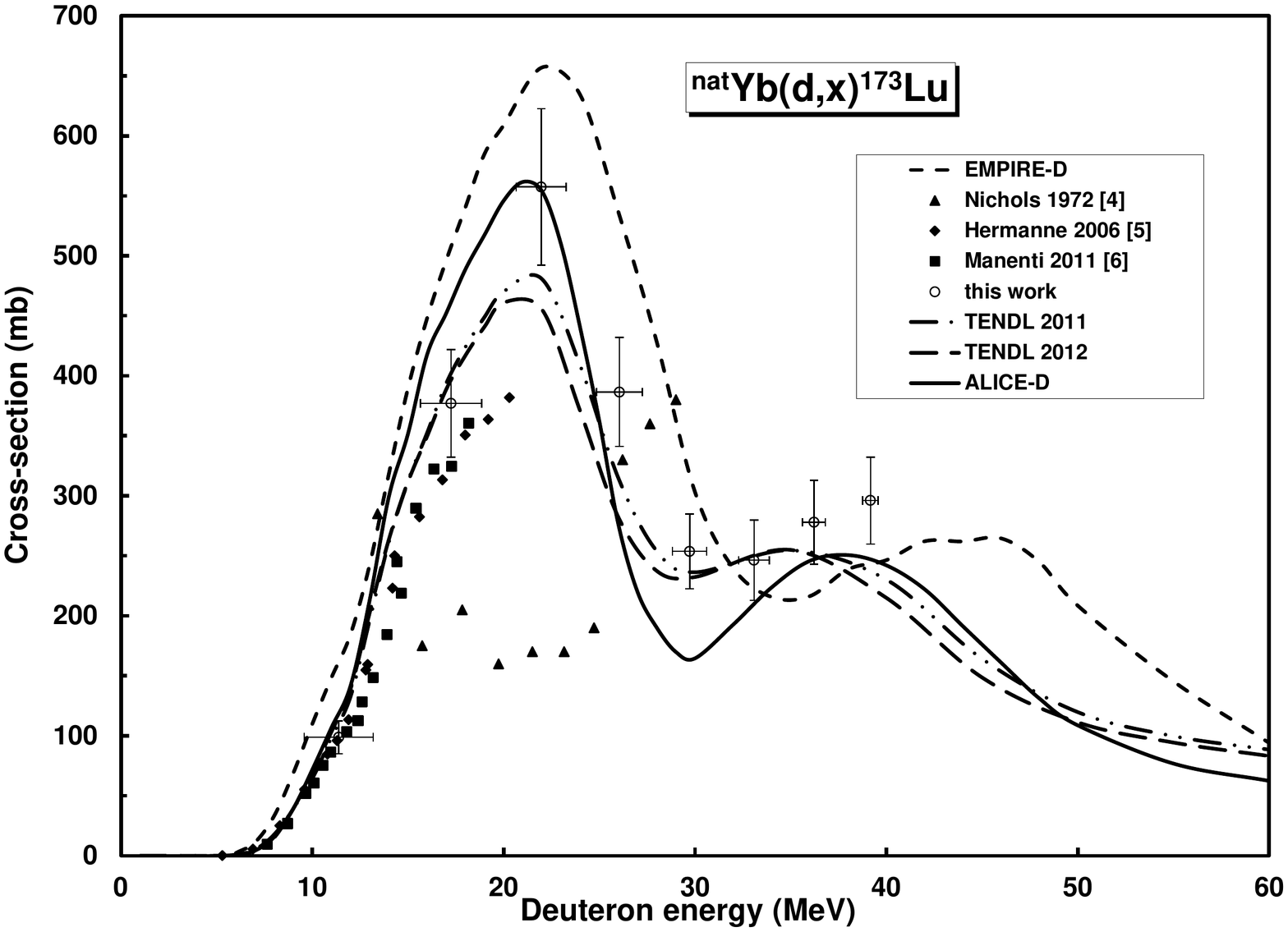}
\caption{Excitation function of the $^{nat}$Yb(d,xn)$^{173}$Lu process}
\end{figure}

\subsubsection{$^{nat}$Yb(d,xn)$^{172}$Lu}
\label{5.1.3}
The acceptable agreement of the experimental and theoretical excitation functions for $^{172}$Lu production is shown in Fig. 4. Reactions on five stable Yb isotopes are contributing (see Table 1). The cross-sections contain the complete contribution of the decay of short-lived isomeric state (3.7 min). The large maximum arises from contribution of the (d,4n) reaction on high abundance $^{174}$Yb. Our new experimental data show good agreement with the earlier results and give new data above 30 MeV. The both TENDL curves run almost together giving good description of the experimental results in the studied energy domain, while ALICE-D shows a larger maximum and EMPIRE-D overestimates above 15 MeV.

\begin{figure}[h]
\includegraphics[scale=0.3]{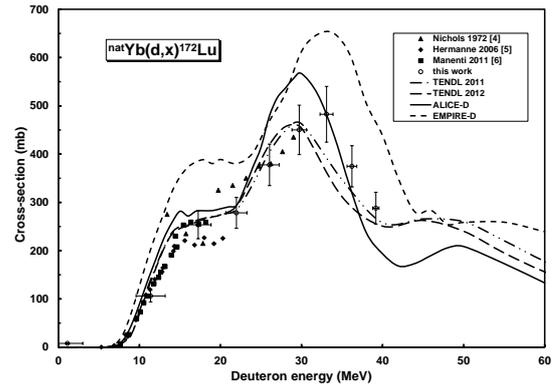}
\caption{Excitation function of the $^{nat}$Yb(d,xn)$^{172}$Lu process}
\end{figure}

\subsubsection{$^{nat}$Yb(d,xn)$^{171}$Lu}
\label{5.1.4}
Below 30 MeV a good agreement of the available experimental data and the theoretical results was obtained (see Fig. 5) for cumulative production of $^{171}$Lu (8.24 d) (including complete decay through isomeric transition of the short-lived (79 s) isomeric state), but our values are systematically higher at higher energies. The three maxima can be connected to the reactions on respectively the low mass ytterbium isotopes, to the $^{174}$Yb and to the $^{176}$Yb target isotopes. Our new data agree well with those of Hermanne \cite{5} and Manenti \cite{6} up to 20 MeV and significantly larger than Nichols' results above 20 MeV. The local maxima era also reproduced well. Now the best approximation is given by the ALICE-D code. The TENDL results are also acceptable under 20 MeV. EMPIRE-D overestimates the experimental data in the whole energy region.

\begin{figure}[h]
\includegraphics[scale=0.3]{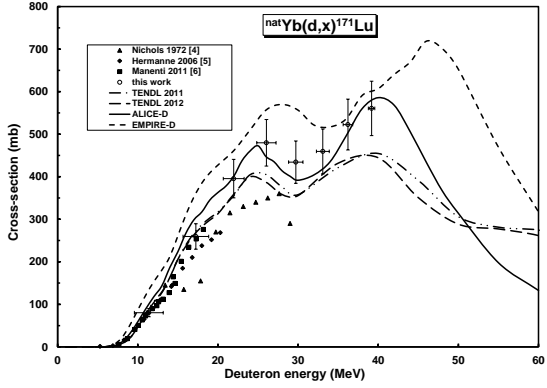}
\caption{Excitation function of the $^{nat}$Yb(d,xn)$^{171}$Lu process}
\end{figure}

\subsubsection{$^{nat}$Yb(d,xn)$^{170}$Lu}
\label{5.1.5}
The excitation function for production of $^{170}$Lu (2.012 d) is shown in Fig. 6. $^{170}$Lu has no isomeric state, so the presented results are direct cross-sections of the ground-state, resulted in from (d,xn) reactions on different Yb stable isotopes (M=170-174). Our new data agree acceptable well with the result of earlier experiments below 20 MeV. The excitation functions predicted in the TENDL show also a good agreement with the experiment. It is difficult to distinguish between the results of the different model calculations, while they mostly run together. Only above 30 MeV is obvious that both TENDL data underestimate the experimental results.

\begin{figure}[h]
\includegraphics[scale=0.3]{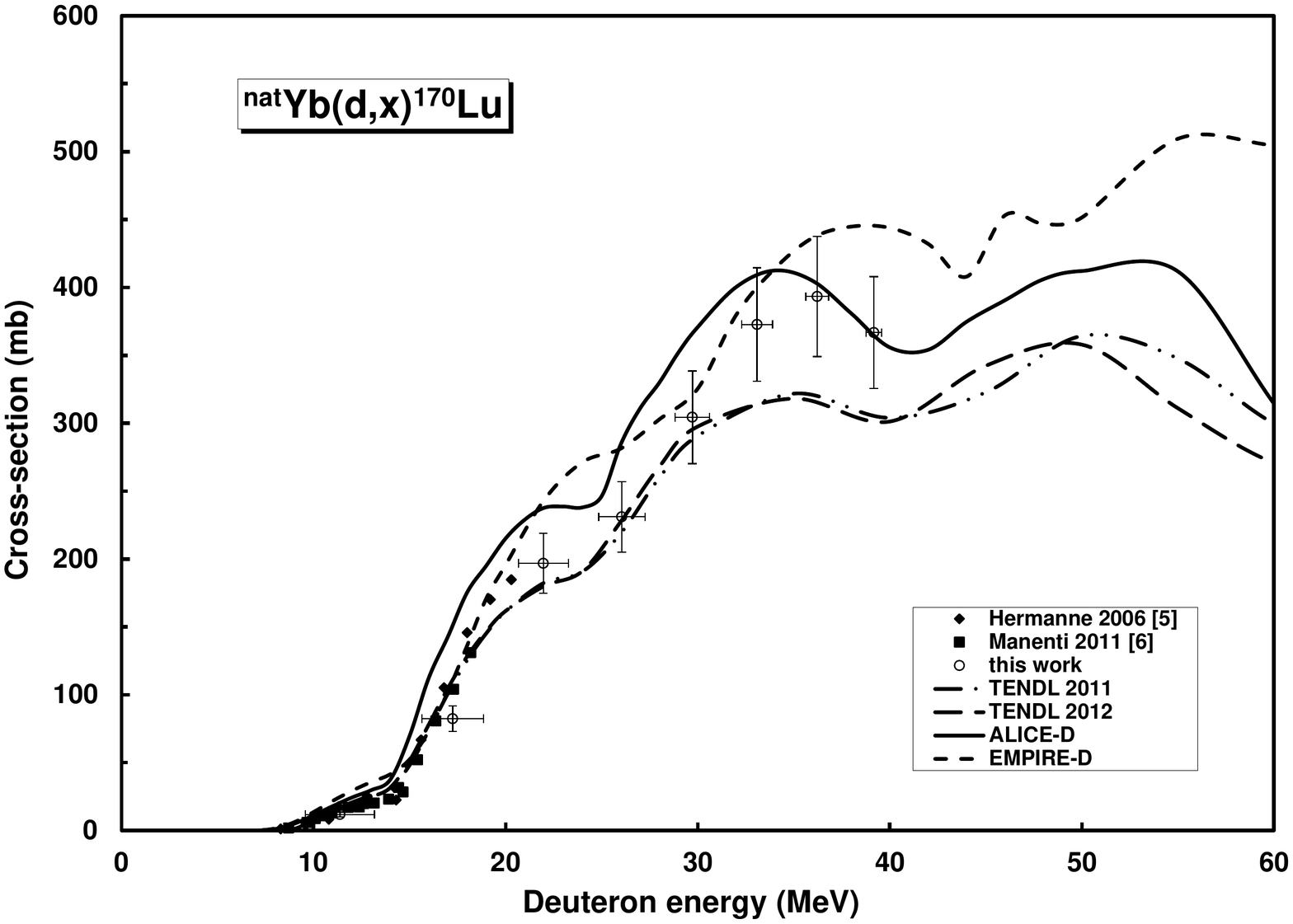}
\caption{Excitation function of the $^{nat}$Yb(d,xn)$^{170}$Lu process}
\end{figure}

\subsubsection{$^{nat}$Yb(d,xn)$^{169}$Lu}
\label{5.1.6}
No earlier experimental data were found for production of $^{169}$Lu (32.018 d), probably due to the high effective reaction threshold not covered in the energy range in \cite{5,6}. The TENDL results reproduce excellently the experimental data (Fig. 7), while both EMPIRE-D and ALICE-D overestimate significantly above 20 MeV. 

\begin{figure}[h]
\includegraphics[scale=0.3]{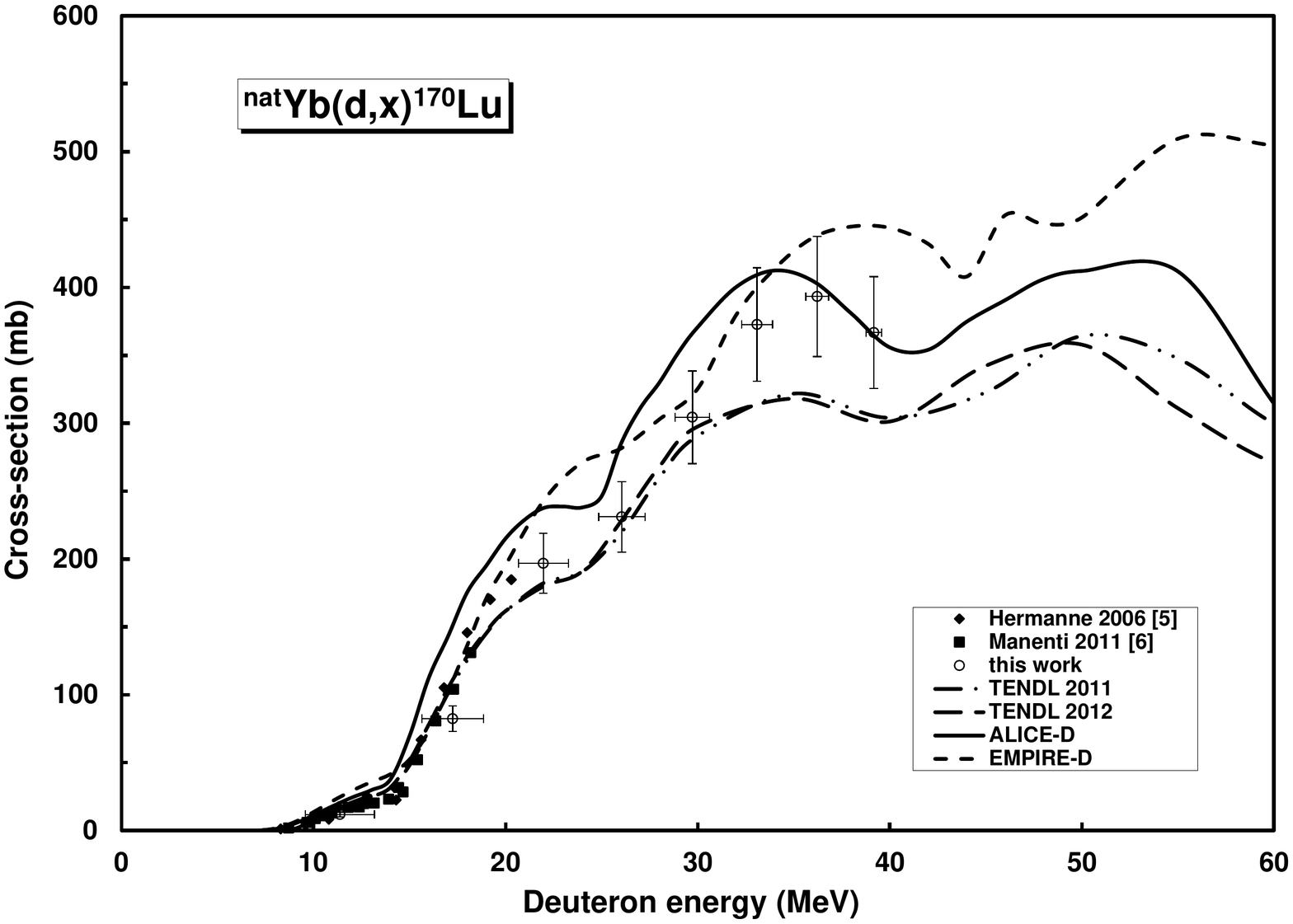}
\caption{Excitation function of the $^{nat}$Yb(d,xn)$^{169}$Lu process}
\end{figure}

\subsection{Cross-sections of ytterbium radioisotopes}
\label{5.2}
The radioisotopes of Yb can be formed through two routes: directly via (d,pxn) reactions and by  decay of simultaneously produced parent radioisotopes of Lu and Tm with the same mass. 

\subsubsection{$^{176}$Yb(d,p)$^{177}$Yb}
\label{5.2.1}
Due to the long waiting time (around 20 h) after EOB before the first measurement we could not identify the  $\gamma$-lines of the $^{177}$Yb (1.911 h) in the spectra. To illustrate the poor predictivity of the model codes for the (d,p) reactions, in Fig. 8 we have compared the experimental data of \cite{5,6} with the TALYS results in the TENDL libraries. The disagreement in the maximum value is around a factor of two. The results of model calculations still significantly underestimate the experimental data, although new improvements for description of (d,p) reactions were implemented in the TALYS code. Phenomenological systematics of the (d,p) cross-sections and the problems of the theoretical descriptions are discussed  in more detail in \cite{25,26}.

\begin{figure}[h]
\includegraphics[scale=0.3]{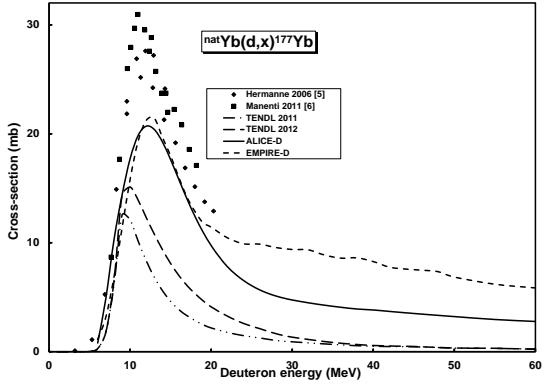}
\caption{Excitation function of the $^{176}$Yb(d,p)$^{177}$Yb process}
\end{figure}

\subsubsection{$^{nat}$Yb(d,x)$^{175}$Yb}
\label{5.2.2}
Fig. 9 shows the cross-sections for cumulative production of $^{175}$Yb (4.185 d) via direct (d,pxn) reactions and from decay of $^{175}$Tm (15.2 min) produced by the $^{176}$Yb(d,2pn) process. According to the theory and to the systematics the (d,2pn) contribution is small. Therefore the cumulative cross-sections mostly reflect the direct (d,pxn) cross-sections. Our present data above 20 MeV are new, in the overlapping region show agreement with the data of Hermanne \cite{5}. The earlier results of Manenti \cite{6} are higher. Similar to the previous case the theoretical results strongly underestimate these processes, the form of the excitation function is reproduced by TALYS only.

\begin{figure}[h]
\includegraphics[scale=0.3]{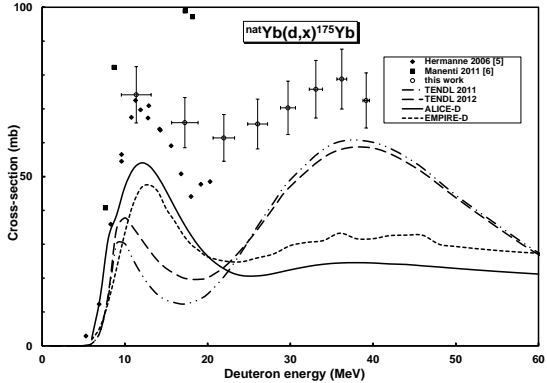}
\caption{Excitation function of the $^{nat}$Yb(d,x)$^{175}$Yb process}
\end{figure}

\subsubsection{$^{nat}$Yb(d,x)$^{169}$Yb}
\label{5.2.3}
The isotope $^{169}$Yb((T$_{1/2}$ = 32.018 d) is obtained through  the direct production via (d,pxn) reactions and is also populated by decay of the simultaneously produced parent $^{169}$Lu (34.06 h). The measured cumulative activation cross-sections are shown in Fig. 10. The agreement with the earlier experimental data and with TALYS, ALICE-D and EMPIRE-D results is acceptable good, although it is difficult to distinguish between the theoretical codes, while they run together up to 25 MeV. Above this energy our new experimental data are better described with ALICE-D and EMPIRE-D. The earlier experimental data of Manenti \cite{6} and Hermanne \cite{5} are in good agreement with our new results in the overlapping energy domain.

\begin{figure}[h]
\includegraphics[scale=0.3]{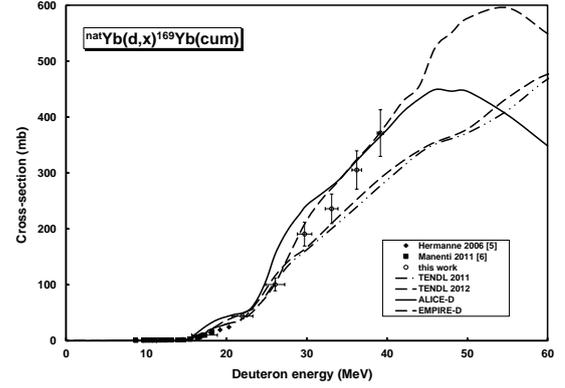}
\caption{Excitation function of the $^{nat}$Yb(d,x)$^{169}$Yb process}
\end{figure}

\subsection{Cross-sections of residual radioisotopes of thulium}
\label{5.3}
The radioisotopes of thulium are produced directly via (d,2pxn) reactions, and/or through decay of simultaneously formed parent Lu and Yb radionuclides. For direct production at low energies the  $\alpha$-particle emission dominates the possible (p,2pxn) (x$\geq$ 2) reaction channels. At higher energies the individual particle emission is more significant. The small contributions from the  $\beta^{-}$-decay of erbium radioisotopes can be neglected due to the low probability of the (d,3pxn) processes.

\subsubsection{$^{nat}$Yb(d,x)$^{173}$Tm}
\label{5.3.1}
According to the Fig. 11 the agreement of the experimental and theoretical data for the directly produced $^{173}$Tm (8.24 h) is not too bad, especially under 30 MeV, taking into account the complex particle emissions for production of this radionuclide. 

\begin{figure}[h]
\includegraphics[scale=0.3]{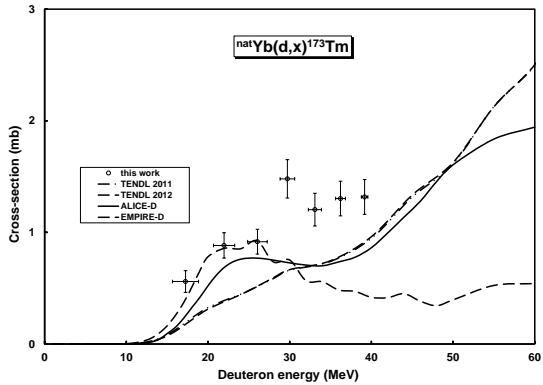}
\caption{Excitation function of the $^{nat}$Yb(d,x)$^{173}$Tm process}
\end{figure}

\subsubsection{$^{nat}$Yb(d,x)$^{168}$Tm}
\label{5.3.2}
The measured excitation function for production of $^{168}$Tm (93.1 d) is shown in Fig. 12. The theory significantly underestimates the experimental values. The both TENDL calculations run together, there is no improvement between the 2011 and 2012 versions. EMPIRE-D and ALICE-D give similar, but still underestimating results. 

\begin{figure}[h]
\includegraphics[scale=0.3]{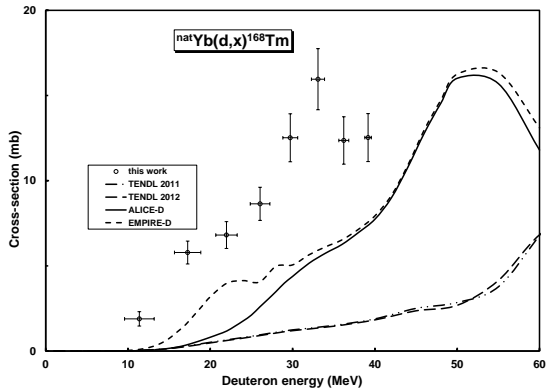}
\caption{Excitation function of the $^{nat}$Yb(d,x)$^{168}$Tm process}
\end{figure}

\subsubsection{$^{nat}$Yb(d,x)$^{167}$Tm}
\label{5.3.3}
The radionuclide $^{167}$Tm (9.25 d) is produced directly and from the decay of short-lived parent $^{167}$Yb (17.5 min). According to Table 1 and to Fig. 13, the threshold for production of the $^{167}$Tm is high in the investigated energy range, but the cross-sections are small. 
We present cross-sections for cumulative production, after complete decay of $^{167}$Lu. The resulted cross-sections contain large uncertainties, especially near the effective threshold, due to the need of correction for interference in the  $\gamma$-ray signal of $^{167}$Tm.
The two measurable $\gamma$-lines of $^{167}$Tm have energies of 207.801(42 \%) and 531.54 keV (1.61\%). As it was mentioned earlier the 207.8 keV line overlaps with the 208.3662 keV  $\gamma$-line of $^{177g}$Lu, which is experimentally practically inseparable because of the energy resolution of the used detector. Therefore we have tried to use the 532 keV  $\gamma$-line of $^{167}$Tm with limited success (low count rates, large scattering). Therefore we have separated the contribution of $^{177g}$Lu in the  $\gamma$-spectra measured 20 day after EOB. It was possible due to the lower  $\gamma$-abundance of the 208 keV  $\gamma$-line of $^{177}$Lu, due to the shorter half-life and to the relatively low cross-sections for cumulative production of $^{177g}$Lu on natural Yb. The experimental excitation function for $^{nat}$Yb(d,xn)$^{177}$Lu(cum) (discussed in section 5.1.1) is also shown in Fig. 13. It is difficult to distinguish, which model code gives the better results, but at least at lower energies under 35 MeV the approximations of EMPIRE-D and ALICE-D are better, but above this energy the experimental values reach the TENDL 2012 curve.

\begin{figure}[h]
\includegraphics[scale=0.3]{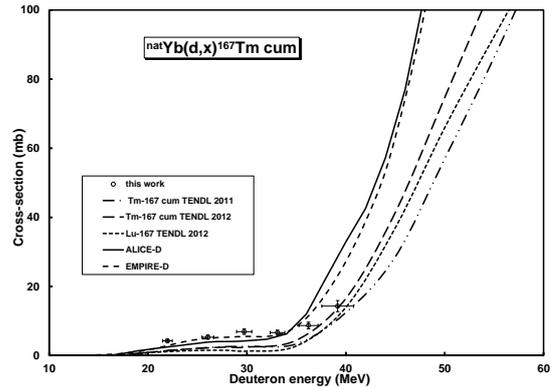}
\caption{Excitation function of the $^{nat}$Yb(d,x)$^{167}$Tm process}
\end{figure}

\subsubsection{$^{nat}$Yb(d,x)$^{165}$Tm}
\label{5.3.4}
The $^{165}$Tm (30.06 h) is produced both directly through $^{nat}$Yb(d,pxn) reactions and indirectly from decay of short lived parent $^{165}$Yb (9.9 min). The effective threshold is close to our maximum energy. The few measured cross-section points are shown in Fig. 14.

\begin{figure}[h]
\includegraphics[scale=0.3]{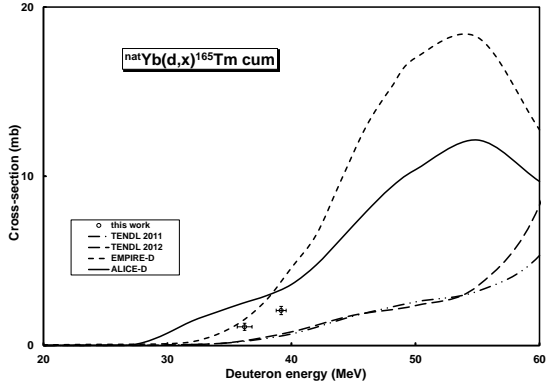}
\caption{Excitation function of the $^{nat}$Yb(d,x)$^{165}$Tm process}
\end{figure}

\begin{table*}[ht]
\tiny
%\small
\caption{Measured cross-sections of the $^{nat}$Yb(d,xn)$^{177,173,172mg,171mg,170,169}$Lu reactions}
\centering
\begin{center}
\begin{tabular}{|p{0.3in}|p{0.3in}|p{0.3in}|p{0.3in}|p{0.3in}|p{0.3in}|p{0.3in}|p{0.3in}|p{0.3in}|p{0.3in}|p{0.3in}|p{0.3in}|p{0.3in}|p{0.3in}|} \hline 
\multicolumn{2}{|p{0.6in}|}{Energy $\pm\Delta$E (MeV)} & \multicolumn{12}{|c|}{Cross-section $\pm\Delta\sigma$ (mb)} \\ \cline{3-14} 
\multicolumn{2}{|p{0.6in}|}{} & \multicolumn{2}{|p{0.6in}|}{${}^{177}$Lu} & \multicolumn{2}{|p{0.6in}|}{${}^{173}$Lu} & \multicolumn{2}{|p{0.6in}|}{${}^{172g}$Lu} & \multicolumn{2}{|p{0.6in}|}{${}^{171g}$Lu} & \multicolumn{2}{|p{0.6in}|}{${}^{170}$Lu} & \multicolumn{2}{|p{0.6in}|}{${}^{16}$${}^{9}$Lu} \\ \hline 
39.2 & 0.4 & 52.0 & 10.8 & 296.1 & 36.2 & 288.0 & 33.0 & 560.6 & 63.9 & 366.8 & 41.2 & 242.2 & 27.8 \\ \hline 
36.2 & 0.6 & 47.6 & 12.6 & 277.9 & 34.9 & 374.7 & 42.7 & 522.6 & 59.7 & 393.4 & 44.2 & 209.4 & 24.1 \\ \hline 
33.1 & 0.8 & 51.8 & 14.0 & 246.4 & 33.4 & 482.5 & 57.7 & 459.6 & 52.7 & 372.7 & 41.8 & 174.5 & 20.3 \\ \hline 
29.7 & 0.9 & 54.1 & 15.5 & 253.7 & 31.2 & 450.2 & 51.2 & 434.3 & 49.8 & 304.4 & 34.2 & 127.9 & 15.0 \\ \hline 
26.0 & 1.2 & 71.0 & 24.9 & 386.5 & 45.3 & 377.7 & 43.0 & 479.8 & 54.7 & 231.0 & 25.9 & 68.4 & 8.4 \\ \hline 
22.0 & 1.3 & 79.4 & 23.0 & 557.3 & 65.1 & 278.3 & 31.9 & 395.5 & 45.2 & 196.8 & 22.1 & 29.7 & 4.2 \\ \hline 
17.2 & 1.6 & 182.7 & 59.2 & 377.1 & 44.8 & 253.5 & 29.0 & 259.4 & 30.0 & 82.4 & 9.3 & 6.7 & 1.6 \\ \hline 
11.4 & 1.8 & 286.3 & 43.7 & 98.9 & 13.7 & 106.1 & 12.2 & 80.4 & 9.5 & 11.8 & 1.7 &  &  \\ \hline 
\end{tabular}

\end{center}
\end{table*}

\begin{table*}[ht]
\tiny
%\small
\caption{Measured cross-sections of the $^{nat}$Yb(d,x)$^{175,169}$Yb and $^{nat}$Yb(d,x)$^{173,172,168,167,165}$Tm reactions}
\centering
\begin{center}
\begin{tabular}{|p{0.3in}|p{0.3in}|p{0.3in}|p{0.3in}|p{0.3in}|p{0.3in}|p{0.3in}|p{0.3in}|p{0.3in}|p{0.3in}|p{0.3in}|p{0.3in}|p{0.3in}|p{0.3in}|} \hline 
\multicolumn{2}{|p{0.6in}|}{Energy $\pm\Delta$E (MeV)} & \multicolumn{12}{|c|}{Cross-section $\pm\Delta\sigma$ (mb)} \\ \cline{3-14} 
\multicolumn{2}{|p{0.6in}|}{} & \multicolumn{2}{|p{0.6in}|}{$^{175}$Yb} & \multicolumn{2}{|p{0.6in}|}{$^{169}$Yb} & \multicolumn{2}{|p{0.6in}|}{$^{173}$Tm} & \multicolumn{2}{|p{0.6in}|}{$^{168}$Tm${}^{  }$} & \multicolumn{2}{|p{0.6in}|}{$^{167}$Tm} & \multicolumn{2}{|p{0.6in}|}{$^{165}$Tm} \\ \hline 
39.2  & 0.4 & 72.5 & 8.2 & 371.4 & 41.7 & 1.3 & 0.2 & 12.5 & 1.4 & 14.3 & 1.6 & 2.1 & 0.2 \\ \hline 
36.2  & 0.6 & 78.8 & 8.9 & 305.4 & 34.3 & 1.3 & 0.2 & 12.4 & 1.4 & 8.7 & 1.0 & 1.1 & 0.2 \\ \hline 
33.1  & 0.8 & 75.8 & 8.5 & 236.0 & 26.5 & 1.2 & 0.2 & 16.0 & 1.8 & 6.6 & 0.7 &  &  \\ \hline 
29.7  & 0.9 & 70.3 & 7.9 & 190.7 & 21.4 & 1.5 & 0.2 & 12.5 & 1.4 & 6.9 & 0.8 &  &  \\ \hline 
26.0  & 1.2 & 65.5 & 7.4 & 100.5 & 11.3 & 0.9 & 0.1 & 8.6 & 1.0 & 5.3 & 0.6 &  &  \\ \hline 
22.0  & 1.3 & 61.4 & 6.9 & 44.0 & 4.9 & 0.9 & 0.1 & 6.8 & 0.8 & 4.3 & 0.5 &  &  \\ \hline 
17.2  & 1.6 & 65.9 & 7.4 & 10.0 & 1.1 & 0.6 & 0.1 & 5.8 & 0.7 &  &  &  &  \\ \hline 
11.4  & 1.8 & 74.2 & 8.3 & 1.5 & 0.3 &  &  & 1.9 & 0.4 &  &  &  &  \\ \hline 
\end{tabular}

\end{center}
\end{table*}

\section{Integral yields}
\label{6}
On the basis of our new experimental cross-section data and the results of the model calculations we determined integral yields for production of the investigated reaction products. The results are so called physical yields \cite{27} corresponding to instantaneous short irradiation. The calculated yields for $^{nat}$Yb(d,xn)$^{177,173,172mg,171mg,170,169}$Lu,  $^{nat}$Yb(d,x)$^{175,169}$Yb and $^{nat}$Yb(d,x)$^{173,172,168,167}$Tm are shown in Fig. 15-16 in comparison with the few experimental integral thick target yield data found in the literature. Dmitriev et al. reported radioactive thick target nuclide yields at 22 MeV incident deuteron energy \cite{7} for the $^{nat}$Yb(d,x)$^{173}$Lu and $^{nat}$Yb(d,x)$^{174g}$Lu reactions. Our calculated yield curve for $^{173}$Lu runs above the data point of Dmitriev at 22 MeV.

\begin{figure}[h]
\includegraphics[scale=0.3]{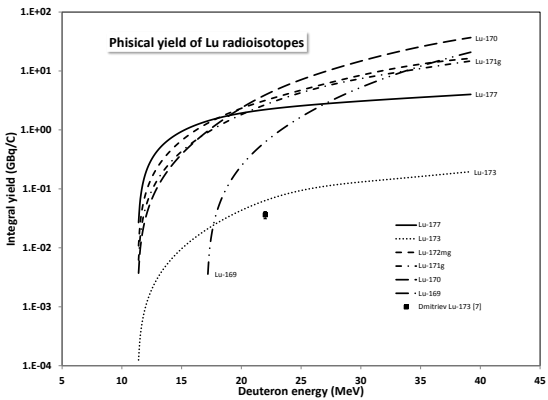}
\caption{Integral yields for production of Lu radioisotopes}
\end{figure}

\begin{figure}[h]
\includegraphics[scale=0.3]{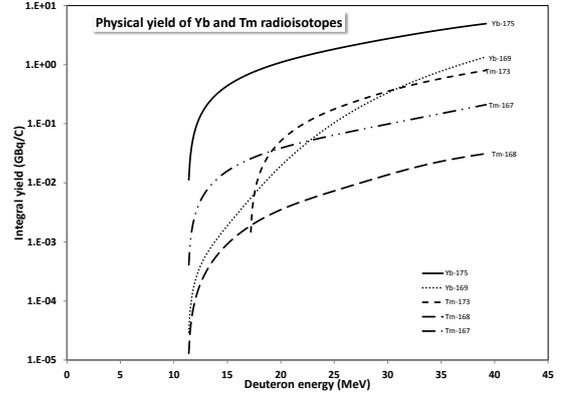}
\caption{Integral yields for production of Yb and Tm radioisotopes}
\end{figure}

\section{Summary and conclusion}
\label{7}
Twelve excitation functions of deuteron induced nuclear reactions on Yb were measured up to 40 MeV. All data above 20 MeV, except for production of $^{171,172,173}$Lu were measured for the first time. The experimental data were compared with the results of EMPIRE-D and ALICE-D calculations and data in the TENDL 2011 and 2012 libraries based on TALYS results. The theoretical calculations describe the shape and the absolute values of the experimental data only with moderate success, and the (d,p) reaction is especially significantly underestimated by the theory. Further improvement of the used model codes is required. The comparison with the low energy earlier experimental data shows excellent agreement.
The obtained experimental data provide a basis for improved model calculations and for different applications in the field of nuclear medicine and for estimating production yields of long-lived products for dose estimations and waste handling. 
Presently the ytterbium has very few industrial uses, but the applications are growing (in laser technology, fiber optic technologies, etc.).
Among the radioactive nuclear reaction products $^{177}$Lu \cite{28,29,30,31}, $^{172}$Lu \cite{32,33,34}, $^{169}$Yb \cite{35}, $^{175}$Yb \cite{36}  and $^{167}$Tm \cite{37} have potential relevance in nuclear medicine for cancer treatment and as tracer in nuclear biology (for a review see \cite{38}).
In the domain of industrial applications, the  $\gamma$-emitter  $^{169}$Yb (39 d half-life) has been used in sealed sources for industrial applications, for substitution of $^{192}$Ir sources applied in conventional non-destructive testing (NDT) projectors, providing better quality of the radiographic pattern \cite{39}.
The longer lived $^{173,174}$Lu are used as radioactive tracers in different processes \cite{40,41} and the  $^{172}$Lu($^{172}$Yb) decay is used for investigation ion implantation in semiconductors by using PAC technique \cite{42},  and for radiotracer studies of oil pipeline flow rates, refinery column residence times, and the performance of a coal liquefaction pilot plant \cite{43}.
Deuteron induced reactions on ytterbium however, are not the most favorable reactions for production of some of these radionuclides as for instance $^{169}$Yb, $^{177}$Lu and $^{175}$Yb can be produced much more economically at reactors via (n,$\gamma$ ) reactions. The optimal routes for the production of $^{167}$Tm are the $^{167}$Er(p,n) \cite{44} and $^{167}$Er(d,2n) \cite{23}.

\section{Acknowledgements}
\label{8}

This work was done in the frame of MTA-JSPS and MTA-FWO (Vlaanderen) research projects. The authors acknowledge the support of research projects and of their respective institutions in providing the materials and the facilities for this work. 
%\FloatBarrier
 


%% References

\clearpage
\bibliographystyle{elsarticle-num}
\bibliography{Ybd}






% \end{thebibliography}

\end{document}